\documentclass[aps,prc,twocolumn,groupedaddress,showpacs]{revtex4}
\usepackage{epsfig}

\begin{document}
\newcommand{\goo}{\,\raisebox{-.5ex}{$\stackrel{>}{\scriptstyle\sim}$}\,}
\newcommand{\loo}{\,\raisebox{-.5ex}{$\stackrel{<}{\scriptstyle\sim}$}\,}

%
%
%
%
%

\title{Formation of hot heavy nuclei in supernova explosions}

\author{A.S.~Botvina$^{1,2}$, and I.N.~Mishustin$^{3,4,5}$}

\affiliation{
$^{1}$Cyclotron Institute, Texas A\&M University, College Station, 
TX 77843, USA\\
$^2$Institute for Nuclear Research, 117312 Moscow, Russia\\
$^3$Institut f\"ur Theoretische Physik, Goethe Universit\"at, 
D-60054 Frankfurt/Main, Germany\\
$^4$Niels Bohr Institute, DK-2100 Copenhagen, Denmark\\
$^5$Kurchatov Institute, RRC, 123182 Moscow, Russia\\
}

\date{\today}

\begin{abstract}
We point out that during the supernova II type explosion the 
thermodynamical condition of stellar matter between the 
protoneutron star and the shock front corresponds to the 
nuclear liquid-gas phase coexistence region, which can be 
investigated in nuclear multifragmentation reactions. 
We have demonstrated, that neutron-rich hot heavy nuclei 
can be produced in this region. The production of these 
nuclei may influence dynamics of the explosion and contribute 
to the synthesis of heavy elements. 
\end{abstract}


\maketitle

In recent years significant progress has been made by nuclear 
community in understanding properties of highly excited nuclear 
systems. Such systems are routinely produced now in nuclear 
reactions induced by hadrons and heavy ions of various energies. 
Under certain conditions, which are well studied experimentally, 
the intermediate nuclear system is produced in a state close to 
statistical equilibrium. At low excitation energies this is nothing 
but a well known compound nucleus. At excitation energies exceeding 
3 MeV per nucleon the intermediate system first expands and then splits into 
an ensemble of hot nuclear fragments (multifragmentation). At excitation 
energies above 10 MeV per nucleon the equilibrated system is composed 
of nucleons and lightest clusters (vaporisation). These 
different intermediate states can be understood as a manifestation 
of the liquid-gas type phase transition in finite nuclear systems 
\cite{Sie}. 
A very good description of such systems is obtained within the framework 
of Statistical Multifragmentation Model (SMM) \cite{SMM}. 

According to present understanding, based on numerous theoretical and 
experimental studies of multifragmentation reactions, prior to the 
break-up a transient state of nuclear matter is formed, where hot nuclear 
fragments exist in equilibrium with free nucleons. This state is 
characterized by a certain temperature $T \sim 3-6$  MeV 
and a density which is typically 
3-5 times smaller than the nuclear saturation density, $\rho_0 \approx 0.15$ 
fm$^{-3}$. Theoretical calculations \cite{SMM} show that relatively heavy 
fragments may survive in the liquid-gas coexistence region. In 
thermodynamic limit these heavy fragments would correspond to the 
infinite liquid phase \cite{Bug}. 
This statistical picture of multifragmentation is confirmed by numerous 
experimental observations, such as an evolution of the fragment mass 
distribution with energy (temperature), fragment correlations revealing 
the critical behavior, anomaly in the caloric curve, 
and by many others (see e.g. \cite{SMM,Bot95,Dag,Schar}). 
Recent experiments (e.g. \cite{GANIL}) directly 
confirm that primary fragments are really hot, their internal
excitation energy may reach up to 3 MeV per nucleon. 
Properties of these hot nuclei can be extracted from multifragmentation 
reactions and used in analyzes of other physical processes, where similar 
nuclei expected to be produced. 

In this work we are going to use the knowledge accumulated 
from multifragmentation 
studies for better understanding nuclear physics associated with 
collapse of massive stars and supernova type II explosions. 
More specifically, we consider the possibility of producing hot heavy 
nuclei in a protoneutron star, and in hot bubble between the protoneutron 
star and the shock front \cite{Jan}. This region is crucial for success 
(or failure) of the supernova explosion. The presence of hot nuclei will 
influence many processes. For example, the electron capture on nuclei plays 
an important role in supernova dynamics \cite{Hix}. 
In particular, the electron capture 
rates are sensitive to the nuclear composition and details of 
nuclear structure (see e.g. \cite{LMP}). 
The neutrino- induced reactions 
are very sensitive to the nuclear structure effects 
and properties of weak interactions in nuclei (see e.g. \cite{Hor}). 
It is also important that the presence of nuclei favors the explosion 
via the energy balance of matter in the bubble \cite{Bethe}.

In the supernova environment, as compared to the 
nuclear reactions, several new important ingredients should be taken 
into consideration. First, the matter at stellar scales must be 
electrically neutral and therefore electrons should be included to 
balance positive nuclear charge. 
Second, energetic photons present in hot matter may change 
nuclear composition via photonuclear reactions.  
And third, the matter is irradiated by a strong neutrino wind from the 
protoneutron star. 
Below we apply a grandcanonical version \cite{Bot} of the SMM to calculate 
mass and charge distributions of nuclear species under these conditions. 

We consider macroscopic volumes of matter consisting of various nuclear 
species $(A,Z)$, nucleons $(n=(1,0)$ and $p=(1,1))$, electrons $(e^-)$
and  positrons $(e^+)$ under condition of electric neutrality. 
In supernova matter there exist several reaction types
responsible for the chemical composition. 
At low densities the most important ones are: 
1) neutron capture and photodisintegration of nuclei
\begin{equation}
(A,Z)+n\rightarrow (A+1,Z)+\gamma~,~
(A,Z)+\gamma\rightarrow(A-1, Z)+n~,
\end{equation}
2) neutron and proton emission by hot nuclei
\begin{equation}
(A,Z)\rightarrow(A-1,T)+n~,~ (A,Z)\rightarrow(A-1,Z-1)+p~,
\end{equation}
and 3) weak processes 
induced by electrons/positrons and 
neutrinos/antineutrinos
\begin{equation} \label{enu}
(A,Z)+e^-\leftrightarrow (A,Z-1)+\nu~,
~(A,Z)+e^+\leftrightarrow (A,Z+1)+\tilde{\nu}~.
\end{equation}
The characteristic reaction times for neutron capture,
photodisintegration of nuclei and nucleon emission can be written as
\begin{eqnarray} 
\tau_{\rm cap}=
\left[\langle\sigma_{nA}v_{nA}\rangle \rho_n\right]^{-1}~,\nonumber\\ 
\tau_{\gamma A}=
\left[\langle\sigma_{\gamma A}v_{\gamma A}\rangle \rho_\gamma\right]^{-1}~,
\nonumber\\
\tau_{n,p}=\hbar/\Gamma_{n,p}~,\nonumber
\end{eqnarray} 
respectively. Here $\sigma_{nA}$ and $\sigma_{\gamma A}$ are the
corresponding cross sections, $v_{nA}$ and $v_{\gamma A}$ are the relative
(invariant) velocities, and $\Gamma_{n,p}$ is the neutron (proton) 
decay width.
Our estimates show that at temperatures and densities of
interest these reaction times vary within the range from 10 to 10$^6$ 
fm/c, that is indeed very short compared to the characteristic
hydrodynamic time of a supernova explosion, about 100 ms \cite{Jan}. 
The nuclear statistical equilibrium is a reasonable assumption  
under these conditions. However, one should specify what kind 
of equilibrium is expected.
For densities $\rho > 10^{-5}\rho_0$ and for the 
expected temperatures of the environment, $T \loo 5$ MeV, we obtain 
$\tau_{\gamma A} >> \tau_{\rm cap}, \tau_{n,p}$, i.e. the 
photodisintegration is less important than other processes. 
There exists a range of densities and 
temperatures, for example, $\rho \goo 10^{-5}\rho_0$ at $T=1$ MeV, 
and $\rho \goo 10^{-3}\rho_0$ at $T=3$ MeV, where the neutron 
capture dominates, i.e. $\tau_{\rm cap} < \tau_{n,p}$. Under 
these conditions new channels for production and decay of 
nuclei will appear (e.g. a fast break-up with emission of 
$\alpha$-particles or heavier clusters) which restore 
the detailed balance. We expect that in this situation an  ensemble of 
various nuclear species will be generated like in a liquid-gas 
coexistence region, as observed in the 
multifragmentation reactions. Here the nuclear system is characterized by the 
temperature T, baryon density $\rho_B$ and electron fraction $Y_e$ (i.e. the 
ratio of the electron density to the baryon ones). 
One may expect that new nuclear effects come into force in this environment. 
For example, at high temperature the masses and level structure in 
hot nuclei can be different from those observed in cold nuclei 
(see e.g. \cite{Igna}). 

The weak interaction reactions are much slower. The direct and inverse 
reactions in eq. (\ref{enu}) involve both free nucleons and all nuclei 
present in the matter. 
It is most likely that at early stages of a supernova explosion 
neutrinos/antineutrinos are trapped inside 
the neutrinosphere around a protoneutron star \cite{Pro}. 
In this case we include the lepton number conservation condition by 
fixing the lepton fraction $Y_L$. 
Out of the surface of the neutrinosphere one should take into 
account the continuous neutrino flux propagating through a hot 
bubble. Due to large uncertainties in the weak interaction rates, 
below we consider three physically distinctive situations: 
1) fixed lepton fraction $Y_L$ corresponding to a $\beta$-equilibrium 
with trapped neutrinos inside the neutrinosphere (early stage); 
2) fixed electron fraction $Y_e$ but without $\beta$-equilibrium inside a 
hot bubble (early and intermediate times); 
3) full $\beta$-equilibrium without neutrino (late times, after cooling 
and neutrino escape). 
The second case corresponds to a non-equilibrium situation which 
may take place in the bubble at early times, 
before the electron capture becomes efficient.

Chemical potential of a species $i$ with baryon number $B_i$, charge $Q_i$
and lepton number $L_i$, which participates in chemical equilibrium, 
can be found from the general expression:
\begin{equation}
\mu_i=B_i\mu_B+Q_i\mu_Q+L_i\mu_L
\end{equation}
where $\mu_B$, $\mu_Q$ and $\mu_L$ are three independent chemical
potentials which are determined from the conservation of total baryon 
number $B=\sum_iB_i$ electric charge $Q=\sum_iQ_i$ and lepton number 
$L=\sum_iL_i$ of the system. This gives
\begin{equation}
 \begin{array}{ll}
\mu_{AZ}=A\mu_B+Z\mu_Q~,~\\
\mu_{e^-}=-\mu_{e^+}=-\mu_Q+\mu_L~,~\\ 
\mu_\nu=-\mu_{\tilde{\nu}}=\mu_L~. 
 \end{array}
\end{equation}
These relations are also valid for nucleons, $\mu_n=\mu_B$ and 
$\mu_p=\mu_B+\mu_Q$. 
If $\nu$ and $\overline{\nu}$ escape freely from the system,
the lepton number conservation is irrelevant and $\mu_L=0$. 
In this case two remaining
chemical potentials are determined from the conditions of baryon 
number conservation and electro-neutrality:
\begin{eqnarray}
\rho_B=\frac{B}{V}=\sum_{AZ}A\rho_{AZ}~,~ 
\rho_Q=\frac{Q}{V}=\sum_{AZ}Z\rho_{AZ}-\rho_e=0~.\nonumber
\end{eqnarray}
Here $\rho_e=\rho_{e^-}-\rho_{e^+}$ is the net electron density. 
The pressure of the relativistic electron-positron gas can be
written as
\begin{eqnarray}
P_e=\frac{\mu_e^4}{12\pi^2}[1+2\left(\frac{\pi
T}{\mu_e}\right)^2+\frac{7}{15}\left(\frac{\pi T}{\mu_e}\right)^4-
 \nonumber\\ 
-\frac{m_e^2}{\mu_e^2}(3+\left(\frac{\pi T}{\mu_e}\right)^2)
] ,\nonumber
\end{eqnarray}
where first order correction due to the finite electron mass is included. 
The net number density $\rho_e$ and entropy density $s_e$ can be
obtained now
from standard thermodynamic relations 
as $\rho_e=\partial P_e/\partial \mu_e$ and $s_e=\partial P_e/\partial T$.
Neutrinos are taken into account in the same way, but as massless particles, 
and with the spin factor twice smaller than the electron one. 

For describing an ensemble of nuclear species under supernova conditions 
one can safely use  the Grand Canonical approximation \cite{SMM,Bot}. 
After integrating out translational degrees of freedom one can write 
pressure of nuclear species as 
\begin{eqnarray} \label{naz}
P_{\rm nuc}=T\sum_{AZ}g_{AZ}\frac{V_f}{V}\frac{A^{3/2}}{\lambda_T^3}
{\rm exp}\left[-\frac{1}{T}\left(F_{AZ}-\mu_{AZ}\right)\right] 
\nonumber\\
\equiv T\sum_{AZ}\rho_{AZ}~,
\end{eqnarray} 
where $\rho_{AZ}$ is the density of nuclear species 
with mass $A$ and charge $Z$.
Here $g_{AZ}$ is the g.-s. degeneracy factor of species $(A,Z)$, 
$\lambda_T=\left(2\pi\hbar^2/m_NT\right)^{1/2}$ is the nucleon
thermal wavelength, $m_N \approx 939$ MeV is the average nucleon mass.
$V$ is the actual volume of the system and $V_f$ is so called
free volume, which accounts for the finite size of nuclear species.
We assume that all nuclei have normal nuclear density
$\rho_0$, so that the proper volume of a nucleus with 
mass $A$ is $A/\rho_0$. At low densities the finite-size effect can be 
included via the excluded volume approximation 
$V_f/V \approx \left(1-\rho_B/\rho_0\right)$. 

The internal excitations of nuclear species $(A,Z)$ play an important 
role in regulating their abundance.  Sometimes they are included through 
the population of nuclear levels known for nearly cold nuclei 
(see e.g. \cite{Jap}). However, in the supernova 
environment not only the excited states but also the binding energies 
of nuclei will be strongly affected by the surrounding matter. 
By this reason, we find it more justified to use another approach 
which can easily be generalized to include in-medium modifications. 
Namely, the internal free energy of species $(A,Z)$ with $A>4$ is 
parameterized in the spirit of the liquid drop model 
\begin{equation}
F_{AZ}(T,\rho_e)=F_{AZ}^B+F_{AZ}^S+F_{AZ}^{\rm sym}+F_{AZ}^C~~,
\end{equation}
where the right hand side contains, respectively, the bulk, 
the surface, the symmetry and the Coulomb terms. The first three terms
are written in the standard form \cite{SMM},
\begin{eqnarray} 
F_{AZ}^B(T)=\left(-w_0-\frac{T^2}{\varepsilon_0}\right)A~~, \\
F_{AZ}^S(T)=\beta_0\left(\frac{T_c^2-T^2}{T_c^2+T^2}\right)^{5/4}A^{2/3}~~,\\
F_{AZ}^{\rm sym}=\gamma \frac{(A-2Z)^2}{A}~~,
\end{eqnarray}
where $w_0=16$ MeV, $\varepsilon_0=16$ MeV, $\beta_0=18$ MeV, $T_c=18$
MeV and $\gamma=25$ MeV are the model parameters which are extracted
from nuclear phenomenology and provide a good description of 
multifragmentation data \cite{SMM,Bot95,Dag,Schar,GANIL}. However, these 
parameters, especially $\gamma$, can be different in hot neutron-rich nuclei, 
and they need more precise determination in nuclear experiments 
(see discussion e.g. in \cite{Bot02}). 
In the Coulomb term we include the 
modification due to the screening effect of electrons. By using the 
Wigner-Seitz approximation it can be expressed as \cite{Lat} 
\begin{eqnarray}
F_{AZ}^C(\rho_e)=\frac{3}{5}c(\rho_e)\frac{(eZ)^2}{r_0A^{1/3}}~~,\\
c(\rho_e)=\left[1-\frac{3}{2}\left(\frac{\rho_e}{\rho_{0p}}\right)^{1/3}
+\frac{1}{2}\left(\frac{\rho_e}{\rho_{0p}}\right)\right]~,\nonumber
\end{eqnarray}
where $r_0=1.17$ fm and $\rho_{0p}=(Z/A)\rho_0$ is the proton 
density inside the nuclei. The screening function $c(\rho_e)$ is 1 
at $\rho_e=0$ and 0 at $\rho_0=\rho_{0p}$. We want to stress that both 
the reduction of the 
surface energy due to the finite temperature and the reduction of the 
Coulomb energy due to the finite electron density favor the formation 
of heavy nuclei. Nucleons and light nuclei $(A \leq 4)$ are considered 
as structureless particles characterized only by mass and proper volume. 

As follows from eq. (\ref{naz}), the fate of heavy nuclei depends 
sensitively on the relationship between $F_{AZ}$ and $\mu_{AZ}$. 
In order to avoid an exponentially divergent contribution to the baryon 
density, at least in the thermodynamic limit ($A \rightarrow \infty$), 
inequality $F_{AZ}\goo \mu_{AZ}$ 
must hold. The equality sign here corresponds to the situation when 
a big nuclear fragment coexists with the gas of smaller clusters \cite{Bug}. 
When $F_{AZ}>\mu_{AZ}$ only small clusters with nearly exponential mass 
spectrum are present. However, there exist thermodynamic conditions 
corresponding to $F_{AZ}\approx\mu_{AZ}$ when the mass distribution of 
nuclear species is broadest. The advantage of our approach is that we 
consider all the fragments present in this transition region, contrary to the 
previous calculations 
\cite{Lat,Latt}, which consider only one ``average'' nucleus characterizing 
the liquid phase. 

Mass distributions and charge-to-mass ratios of nuclear species were 
calculated for three sets of physical
conditions expected in protoneutron stars and in supernova
explosions. We take baryon number $B=$1000 and perform calculations 
for all fragments with 1$\leq A \leq$1000 and 0$\leq Z \leq A$ 
in a box of fixed volume $V$. 
The fragments with larger masses ($A>$1000) can be produced only at 
a very high density $\rho_B \goo 0.5\rho_0$, which is appropriate for 
the regions deep inside the protoneutron star, and which is not 
considered here. 
First we consider the case when lepton fraction 
is fixed as expected inside a neutrinosphere. 
    \begin{figure}[ht]
\hspace{-3mm}
\includegraphics[width=1.1\columnwidth]{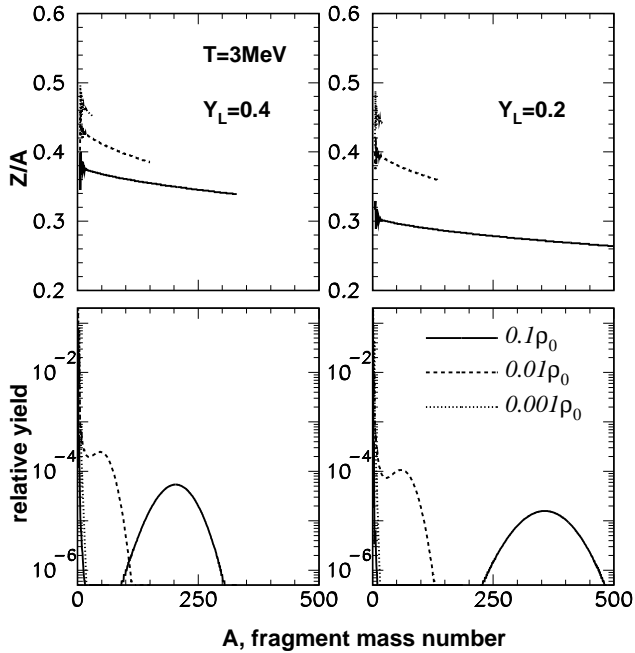} 
    \caption{
Mass distributions (bottom panels) and average 
charge to mass number ratios (top panels) of hot primary fragments produced 
at temperature $T=$3 MeV, and for different densities (in shares 
of the normal nuclear density $\rho_0$) shown on the figure. The calculations 
are for the lepton number conservation: the lepton fraction is 
$Y_L=$0.4 on the left panels, and $Y_L=$0.2 on the right ones. 
}
   \end{figure}
Figure~1 shows the results for 
$Y_L$=0.4 and $Y_L$=0.2. For a typical temperature $T=3$ MeV we find 
that the islands of heavy nuclei, $100<A<500$, 
can appear only at relatively high baryon density, $\rho_B=0.1\rho_0$. 
These nuclei are neutron-rich: $Z/A\approx 0.38$ and $Z/A\approx 0.27$ 
for $Y_L$=0.4 and $Y_L$=0.2, respectively. The $Z/A$ ratios are 
decreasing with $A$ less rapidly than in the nuclear 
multifragmentation case \cite{Bot01}. This 
can be explained by the screening effect of electrons. 
The width of the charge distribution at given $A$ is 
determined by $T$ and $\gamma$: $\sigma_Z \approx \sqrt{AT/8\gamma}$ 
\cite{Bot,Bot01}. At lower density, 
$\rho_B=0.01\rho_0$, the mass distribution is rather flat up to 
$A\approx 80$ and then decreases rapidly for larger $A$. For
$\rho=10^{-3}\rho_0$ only light clusters are present and the mass 
distribution drops exponentially. We have also found that for $T=5$ MeV the 
island of heavy nuclei 
($300<A<700$) is observed at very high densities, 
$\rho_B \approx 0.3\rho_0$,
but at $\rho_B=0.1\rho_0$ the mass distribution is very broad, up to 
$A\approx 180$. This picture is similar to the nuclear liquid-gas 
coexistence region observed in finite systems \cite{SMM}. 

Let us consider the situation more appropriate for a hot bubble at early 
times of a supernova explosion, when 
the  neutrino wind from the core interacts with the infalling matter. 
In this case only the electron fraction is fixed, and 
the electron and proton chemical potentials 
are determined independently, without using the equilibrium relation 
$\mu_e=-\mu_Q$. Corresponding results for $Y_e=0.4$ and  $Y_e=0.2$ at 
$T=1$ MeV and several 
baryon densities are presented in Fig.~2 (left top and bottom panels). 
    \begin{figure}
\hspace{-3mm}
\includegraphics[width=1.1\columnwidth]{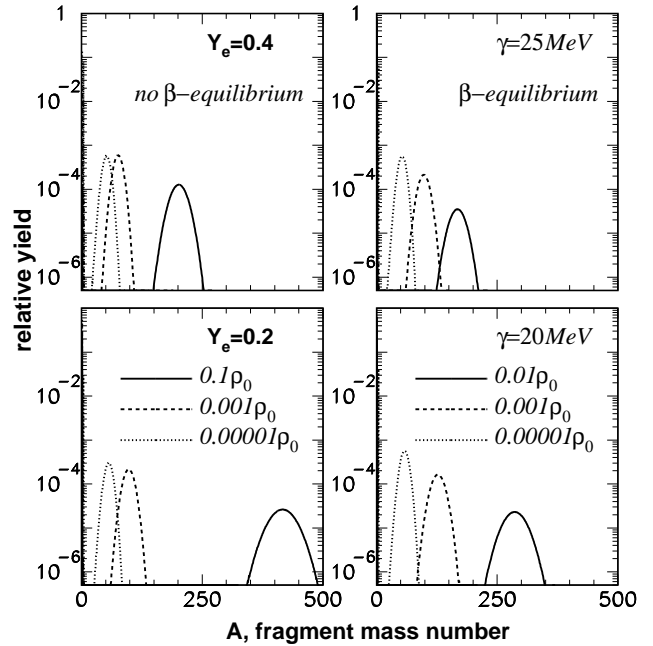} 
    \caption{
Mass distributions of hot primary fragments produced 
at temperature $T=$1 MeV, and for different densities. Left 
panels are for the case of the fixed electron fractions $Y_e$ 
(no $\beta$-equilibrium); right panels are for the full 
$\beta$-equilibrium (without neutrino capture), and for different 
symmetry energy coefficients $\gamma$. 
}
   \end{figure}
One can 
see that heavy nuclei, $50<A<500$, can be produced in very broad range 
of densities, $0.1\rho_0>\rho_B>10^{-5}\rho_0$. At given density the 
mass distribution has a Gaussian shape. In the $Y_e=0.4$ case the most 
probable nuclei, corresponding to the maxims of distributions, 
have $Z/A$ ratios  0.400, 0.406, and 0.439, for densities $0.1\rho_0$, 
$10^{-3}\rho_0$, and $10^{-5}\rho_0$, respectively. 
The Gaussian mass distributions may in some cases justify earlier calculations 
\cite{Lat,Latt}, when only one heavy nuclear species 
was assumed at each density. As seen from the bottom panel, 
changing the electron fraction from 0.4 to 0.2 leads to a significant 
increase of nuclear masses.  Also, the nuclei become more neutron rich: 
the corresponding $Z/A$ ratios are 0.280, 0.359, and 0.420. 
It is interesting
to note that the combination of chemical potentials $\mu_e+\mu_p-\mu_n$ which
drives the electron capture might be quite large in this case. For instance,
for $Y_e=0.4$ this value is 14 MeV at $\rho_B=10^{-3}\rho_0$,   
so that the electron 
capture rate will be significantly enhanced compared to the equilibrium 
rate at $T=1$ MeV\footnote{an enhancement factor of about 20 is obtained 
assuming $E^2$ dependence of the cross section}.  

In the right panels of Fig.~2 we show results for the case 
of full $\beta$-equilibrium. 
We observe that the distributions are quite similar
to the ones shown in the left panels. For the case of the standard 
symmetry energy with $\gamma$=25 MeV we obtain that the $Z/A$ ratios 
of the most probable nuclei are 0.311, 0.359, and 0.435, for densities 
$0.01\rho_0$, $10^{-3}\rho_0$, and $10^{-5}\rho_0$, respectively. 
It is seen from the bottom panel that by slightly decreasing 
the $\gamma$-coefficient in the symmetry energy one can shift mass 
distributions to higher masses. 
The nuclei in this case are even more neutron rich, 
the corresponding $Z/A$ ratios are 0.241, 0.317, and 0.417. 
For easier comparison, in left and right panels we present 
calculations for the baryon densities 
$10^{-3}$ and $10^{-5}\rho_0$, which are more realistic 
in the bubble. The comparison shows that in the course of 
$\beta$-equilibration the mass 
distributions may shift by 10-40 mass units. This shift also characterizes
uncertainties in the nuclear composition associated with the electron capture
reactions. 

Figure~3 shows the fractions of free electrons and neutrons 
as a function of baryon density. 
    \begin{figure}
\hspace{-3mm}
\includegraphics[width=1.1\columnwidth]{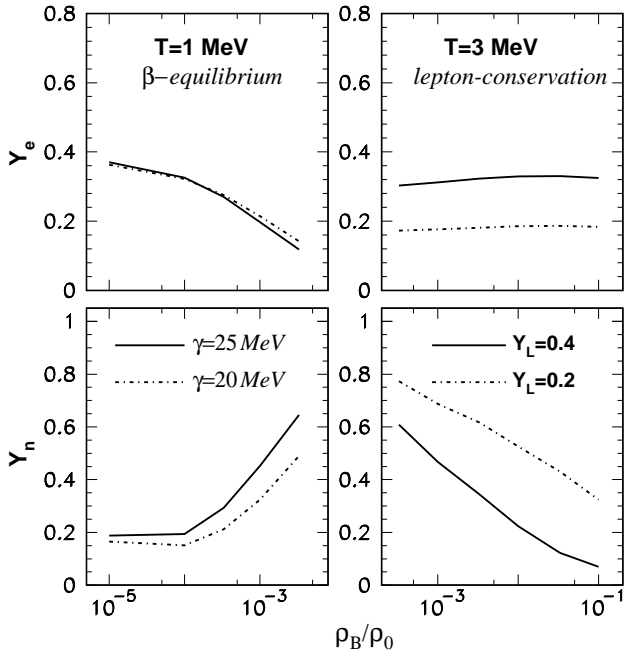} 
    \caption{
Average shares of electrons $Y_e$ (top panels) and 
free neutrons $Y_n$ (bottom panels) versus density. 
Left panels are for the $\beta$-equilibrium with $T=$1 MeV and 
different $\gamma$. 
Right panels are for the lepton conservation with $T=$3 MeV and 
different $Y_L$. 
}
   \end{figure}
On the left panels it was calculated for the 
$\beta$-equilibrium neutrinoless matter at $T=1$ MeV. 
These particles, besides heavy nuclei, dominate in the supernova matter.
As was realized long ago \cite{Bethe}, the 
electrons are absorbed at large densities because of the high 
electron chemical potential. In this work we point out the importance of free 
neutrons in the liquid-gas coexistence region for maintaining a high rate of 
nuclear reactions. 
A noticeable change in the trend is seen at $\rho_B\approx 10^{-4}\rho_0$.
At lower density $Y_n \approx$ 0.2 that means that 80\% of neutrons are 
trapped in nuclei. At higher densities more and more neutrons are dripping
out of nuclei, and at $\rho_B>10^{-3}\rho_0$ more than half of the 
neutrons are free. This behavior correlates with the decreasing number 
of electrons and a relatively small share of heavy nuclei in the system. 
However, at higher densities, the structure of matter 
changes because of the neutrino/antineutrino and electron/positron capture 
reactions \cite{Bethe}.
The case of lepton conservation is shown on the right panels of Fig.~3. 
One can see that in this case the number of free neutrons drops with 
density reflecting formation of very big nuclei and transition to the 
liquid phase at $\rho_B \rightarrow \rho_0$. 

Finally, we find thermodynamic conditions, which allow to produce 
heavy elements known as the solar element abundances. Here we are not 
pretending to fit the fine structure of these yields, in particular, 
the pronounced peaks caused by the structure effects of cold nuclei. 
Our goal is to propose an explanation for the 
gross distribution of the elements. 
    \begin{figure}
\hspace{-3mm}
\includegraphics[width=1.1\columnwidth]{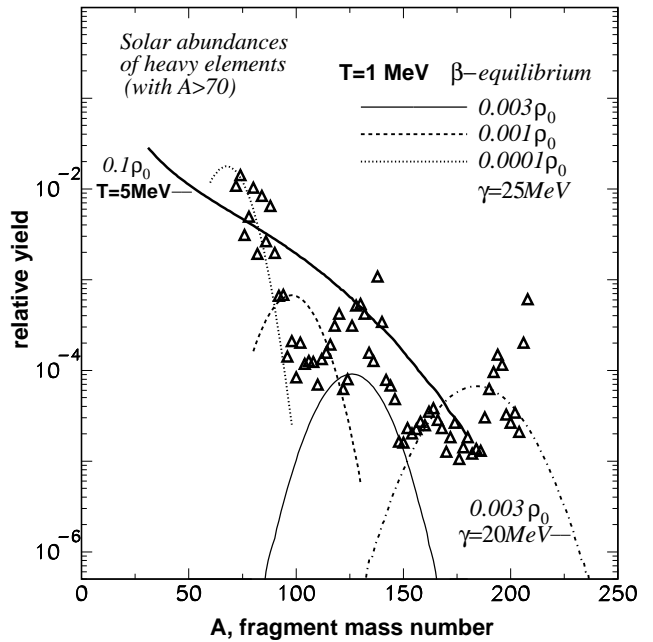} 
    \caption{
The mass distribution of heavy elements (with $A>$70) 
in Solar System. The experimental data (triangles) are taken 
from \cite{Qui}. The lines are calculations at densities shown on 
the figure. The thick solid line is for the lepton conservation at 
$Y_L=$0.4 and $T=$5 MeV. The thin solid, dashed, and dotted lines 
are for the full $\beta$-equilibrium, with the standard 
symmetry energy coefficients $\gamma$, and $T=$1 MeV; 
the dot-dashed line is the same but for the reduced $\gamma$. 
}
   \end{figure}
The results of our fit 
are presented in Fig.~4. It is interesting that by choosing $T=5$ MeV, 
$\rho_B=0.1\rho_0$ and $Y_L=0.4$ we are able to achieve a reasonable overall 
fit. This means that elements heavier than Fe could be produced 
in protoneutron stars. However, it is difficult to find a mechanism for 
ejecting this material from the star. Another possibility is to consider 
lower densities and temperatures expected in the bubble region. In Fig.~4 
we also show the fit with $T=1$ MeV and several baryon densities from 0.003 
to $10^{-4}\rho_0$. In this illustrative calculations we assume that the 
total baryon number is evenly distributed between the regions with different 
densities. A better choice would correspond to a temperature and density 
profile obtained from hydrodynamical simulations of the supernova 
explosion. As expected \cite{Jan}, during the explosion, 
dynamical instabilities (convection and other processes) 
may lead to a large-scale mixing of matter and to large density fluctuations. 
Appearance of regions with higher density favors production of 
heavy elements. 
One can see that we can, in principal, explain the element abundances 
in this way.  Also, a variation of the parameters of hot nuclei, 
in particular, the $\gamma$-coefficient in the symmetry energy, can 
influence the final predictions. 

One should bear in mind that the mass distributions which are presented here 
correspond to hot primary nuclei. After ejection these nuclei will undergo
de-excitation. At typical temperatures considered here ($T\loo 3$ MeV)
the internal excitation 
energies are relatively low, less than 1.0 MeV/nucleon. As well known
from calculations \cite{SMM} and nuclear experiments \cite{Dag,Schar,GANIL}, 
de-excitation of nuclei with $A\leq$ 200 
will go mainly by means of the nucleon emission. 
Then the resulting distributions of cold nuclei are not very different from 
the primary ones, they are shifted to lower masses by several units. One 
should expect that shell effects 
(which, however, may be modified by surrounding electrons) 
will play an important role at the 
de-excitation stage leading to the fine structure of the mass distribution.   
We believe that after the de-excitation of hot nuclei, corresponding to 
the time when the ejected matter reaches very low densities, 
the r-process may be responsible for the final redistribution 
of the element abundances leading to the pronounced peaks 
around $A\approx$80, 130 and 200 \cite{Qui}. 

In conclusion, we have used statistical approach to study production of hot 
heavy nuclei during the collapse of massive stars and subsequent supernova 
explosions. Mass and charge distributions of such nuclei have been calculated
under different assumptions regarding temperature, baryon density, electron 
and lepton 
composition of the matter. We have demonstrated that this mechanism can 
contribute significantly to the production of heavy elements in supernova 
environment and explain gross features of the element abundances. 
We also expect that the production of hot heavy 
nuclei may influence the explosion dynamics through both the energy 
balance and the capture/production of electrons and neutrinos.

This work was supported in part by the Deutsche Forschung 
Gemeinschaft (DFG) under grant 436 RUS 113/711/0-1, the Russian Fond of 
Fundamental Research (RFFR) under grant 03-02-04007 and Russian Ministry of 
Industry, Science and Technology under grant SS-1885.2003.2.1. 
A.S.B. thanks the Cyclotron Institute of Texas A\&M University, 
Indiana University (Bloomington, IN), and Gesellschaft f\"ur 
Schwerionenforschung (Darmstadt, Germany), where parts of this work 
were done, for hospitality and support.


\end{document}